\documentclass[sn-mathphys]{sn-jnl}
\usepackage[utf8]{inputenc}
\usepackage{microtype}
\usepackage{mathtools}
\usepackage{booktabs}
\usepackage{amsfonts}
\usepackage{amsbsy}
\usepackage{physics}
\usepackage{caption}
\usepackage{subcaption}
\usepackage{xcolor}
\usepackage{graphicx}
\usepackage{navigator}
\usepackage{natbib}

\newcommand{\Real}[0]{\mathbb{R}}
\newcommand{\X}[0]{\mathcal{X}}
\newcommand{\Hilbert}[0]{\mathcal{H}}
\newcommand{\vecx}[0]{\mathbf{x}}

\begin{document}

\abstract{
Exploiting the properties of quantum information to the benefit of machine learning models is perhaps the most active field of research in quantum computation. This interest has supported the development of a multitude of software frameworks (e.g. Qiskit, Pennylane, Braket) to implement, simulate, and execute quantum algorithms. Most of them allow us to define quantum circuits, run basic quantum algorithms, and access low-level primitives depending on the hardware such software is supposed to run. For most experiments, these frameworks have to be manually integrated within a larger machine learning software pipeline. The researcher is in charge of knowing different software packages, integrating them through the development of long code scripts, analyzing the results, and generating the plots. Long code often leads to erroneous applications, due to the average number of bugs growing proportional with respect to the program length. Moreover, other researchers will struggle to understand and reproduce the experiment, due to the need to be familiar with all the different software frameworks involved in the code script. We propose QuASK, an open-source quantum machine learning framework written in Python that aids the researcher in performing their experiments, with particular attention to  quantum kernel techniques. QuASK can be used as a command-line tool to download datasets, pre-process them, quantum machine learning routines, analyze and visualize the results. QuASK implements most state-of-the-art algorithms to analyze the data through quantum kernels, with the possibility to use projected kernels, (gradient-descent) trainable quantum kernels, and structure-optimized quantum kernels. Our framework can also be used as a library and integrated into pre-existing software, maximizing code reuse. 
}


\title[Quantum Advantage Seeker with Kernels (QuASK)]{Quantum Advantage Seeker with Kernels (QuASK): a software framework to speed up the research in quantum machine learning}

\author[1]{\fnm{Francesco} \sur{Di Marcantonio}}

\author*[2,4]{\fnm{Massimiliano} \sur{Incudini}}\email{massimiliano.incudini@univr.it}

\author[3,4]{\fnm{Davide} \sur{Tezza}}

\author*[1]{\fnm{Michele} \sur{Grossi}}

\affil[1]{\orgname{European Organization for Nuclear Research (CERN)}, \orgaddress{\city{Geneva}, \postcode{1211}, \country{Switzerland}}}

\affil[2]{\orgdiv{Dipartimento di Informatica}, \orgname{Universit\`a di Verona}, \orgaddress{\city{Verona}, \postcode{34137}, \country{Italy}}}

\affil[3]{\orgdiv{Dipartimento di Matematica}, \orgname{Universit\`a di Trento}, \orgaddress{\city{Povo}, \postcode{38123}, \country{Italy}}}

\affil[4]{\orgname{Data Reply s.r.l.}, \orgaddress{\city{Turin}, \postcode{10126}, \country{Italy}}}

\date{\today}
\maketitle




\section{Introduction}

Breakthroughs in quantum technologies have allowed the construction of small-scale prototypes of quantum computers \cite{madsen2022quantum, dumitrescu2022dynamical, huang2022quantum}, namely NISQ devices \cite{preskill2018quantum}. Even though many sources of noise may corrupt the execution on these devices \cite{pelofske2022quantum}, we are able to run a certain class of algorithms \cite{nisq_algorithms_2022} which compromises the strong theoretical speedup of fault-tolerant quantum algorithms \cite{montanaro2016quantum} to achieve shorter, less noisy computations. A large subset of the NISQ-ready algorithms is dedicated to the development of machine learning models. 

One of the most interesting technique among them are the quantum classifiers \cite{Schuld_2019, havlivcek2019supervised, mengoni2019kernel}: the function $f(x) = \Trace[\rho_\vecx \rho_w]$, where $\rho_\vecx$ represents the encoding of a data point $\vecx$ in a quantum state through the feature map $\ketbra{0}{0} \mapsto U(\vecx)\ketbra{0}{0} = \rho_\vecx$ and $\rho_w$ represents the weight vector encoded through the mapping $\ketbra{0}{0} \mapsto W\ketbra{0}{0} = \rho_w$, can be interpreted as a linear\footnote{The linearity of the model means that the model can be expressed as linear transformation in the Hilbert space of the quantum system (e.g. matrix multiplication for finite-dimensional Hilbert spaces).
However, the classifier is non-linear with respect to the space in which the original data lie, for the effect of the feature map.} model  \cite{schuld2021machine}.
Such a function can be immediately used to solve supervised learning tasks. 
By choosing the weight mapping to be parametric $W(\theta)$, we can train the parameters to minimize some loss function using gradient-descent-based techniques: such an approach is named \emph{quantum neural network} \cite{mitarai2018quantum, abbas2021power}. 
However, the training phase of these models could be affected by barren plateau \cite{mcclean2018barren, holmes2021barren}, \textit{i.e.} the flat loss landscape, where the variance of the gradient vanishing exponentially fast with respect to the number of qubits. 
Highly entangled states \cite{marrero2021entanglement}, noise \cite{wang2021noise}, global measurement \cite{arrasmith2021effect}, and expressibility of the feature map \cite{holmes2022connecting} have been linked to the appearance of barren plateau.
To avoid such a problem, \emph{Quantum Kernel Estimation} (QKE) \cite{Schuld_2019, havlivcek2019supervised, mengoni2019kernel} algorithm can be used in a hybrid form - we implement a quantum kernel function $\kappa(\vecx, \vecx') = \Trace[\rho_\vecx \rho_{\vecx'}]$, quantifying the similarity between two encoded data points, with a classical machine learning algorithm. 
The training of the model is classical and is expected to end successfully\footnote{In some cases the quantum kernel values can concentrate around an average value, requiring a large number of shots to be estimated \cite{kubler2021inductive, thanasilp2022exponential}.} and efficiently due to the representer theorem \cite{scholkopf2001generalized}. 
Classical kernel methods are a cornerstone of machine learning, and have been applied to any sort of task including signal processing \cite{perez2004kernel, rojo2018digital, camps2006kernel}, bioinformatics \cite{camps2006kernel, ben2008support}, and image processing \cite{wang2014pet, yang2001face}.

A clear benefit in using quantum kernel estimation to enhance machine learning applications has still to be found \cite{schuld2022quantum}.
Quantum kernels have been shown to improve the performances of classical machine learning algorithms for some problems, such as the prediction of the output of quantum systems \cite{Huang2021, huang2022advantage}, and in learning from distributions based on the discrete logarithm \cite{liu2021rigorous}. 
They have been applied to several real-world, industrial scale problems such as anomaly detection \cite{nana_anomaly_det_2018}, fraud detection \cite{di2021quantum, grossi2022mixed, kyriienko2022unsupervised}, the effectiveness of pharmaceutical treatments \cite{krunic2022quantum}, and supernova classifications \cite{peters2021machine}.
These approaches have been experimentally tested on superconducting \cite{peters2021machine, wang2021towards}, optical \cite{bartkiewicz2020experimental}, and NMR \cite{kusumoto2021experimental} quantum devices, and their effectiveness is usually assessed empirically.

Most of these experiments share, at least partially, a common structure: dimensionality reduction techniques, used to limit the number of quantum resources needed for the computation; the scaling of the input; the choice of the quantum kernel; the evaluation of the quantum kernel. 
The researcher is usually in charge of developing a software prototype, which requires the knowledge of many different software frameworks and platforms: the ones dealing with the machine learning tasks \cite{NEURIPS2019_9015, chollet2015keras}, and the one dealing with quantum computing \cite{bergholm2018pennylane, Qiskit, broughton2020tensorflow, strawberryfields, Paddlequantum}. 
As the prototype becomes larger, the probability of introducing \emph{bugs} in the code increases \cite{lipow1982number}, possibly leading to erroneous results \cite{fidler2017metaresearch, botvinik2020variability, campos2021qbugs}. Well-organized code has been shown to facilitate code reuse and reproducibility \cite{trisovic2022large, patrick_mineault_2021_5796873}.
Minimizing the quantity of code needed to run an experiment has clear benefit in speeding up the research, reducing the time spent to learn, and put the software in a production environment.

We propose QuASK (Quantum Advantage Seeker with Kernel), a Python3 software framework unifying under a single interface all the features to run experiment with quantum kernels.
QuASK can be run from the terminal using a single command line which specifies how to operate on the given data. 
Within the same command, the researcher can specify to analyze the data and subsequently generate graphics.
QuASK can also be used as a library, to be integrated within an existing pipeline. 
Finally, the open-source nature of the framework allows the user to integrate further capabilities into the software, having them immediately available through the command line interface. 
QuASK is freely available at \url{https://github.com/CERN-IT-INNOVATION/QuASK} and the documentation is available at \url{https://quask.readthedocs.io/en/latest/index.html}.

\section{Theoretical aspects of Quantum Kernels}

A binary, symmetric function $K: \X \times \X \to \Real$ is a \emph{kernel function} if positive definite (pd), \textit{i.e.}
\begin{equation} 
    \sum_{i=1}^n \sum_{j=1}^n c_i c_j K(x_i, x_j) \ge 0
\end{equation}
for all $x_1, ..., x_n \in \X$ given the real\footnote{The above definition of real-valued kernel is true if we require also the kernel matrix to be symmetric.} coefficients $c_1, ..., c_n \in \Real$. Supposing $k$ continuous, we can associate a linear Hilbert-Schmith integral operator
\begin{equation}
    [T g](x') = \int_{x \in \X} K(x', x) g(x) dx.
\end{equation}
whose eigenfunctions $\{ e_i \}_{i=1}^\infty$ form an orthonormal basis of square-integrable functions, and the sequence of corresponding eigenvalues $\{ \lambda_i \}_{i=1}^\infty$ is non-negative. In such a case,
\begin{equation}
    K(x, x') = \sum_{i=1}^\infty \lambda_i e_i(x) e_i(x')
\end{equation}
For $x, x' \in \X \subseteq \Real^d$, some examples of kernels are:
\begin{align}
    K_\text{l}(x, x') & = x \cdot x' && \text{Linear} \\
    K_\text{p}(x, x') & = (x \cdot x' + b)^r && \text{Polynomial of degree }r \\
    K_\text{rbf}(x, x') & = \exp(-\alpha \lVert x - x'\rVert) && \text{RBF or Gaussian } (\alpha > 0) \\
    K_\text{eq}(x, x') & = 1 - \delta_{x, x'} && \text{Equality kernel}
\end{align}
while, for example, $\vert x-x'\vert$ is not a valid kernel due to the lack of positive definiteness.
Important class of kernels are the \emph{translation-invariant kernels} $K(x, x') = \Phi(x - x')$ given that $\X$ is a vector space (the Gaussian kernel is an instance of such a family), and the \emph{group kernel} $K(x, x') = \Phi(x^{-1} x')$ given $\X$ has a group structure. 
Kernels can form other kernels: a non-negative linear combination of kernels, a product, and the limit of a  kernel sequence (if exists) are kernels too.
$K_\text{eq}$ is a valid non-continuous kernel, which can be obtained as the limit for $n \to \infty$ of $\exp\{-n\lVert x - x' \rVert\}$. 
A larger list of kernels and their compositions can be found in \cite{gpbook, duvenaud2014automatic}.

Positive definite kernels can be thought as a generalization of the notion of inner product due to the strong relationship with the concept of \emph{Reproducing Kernel Hilbert Space} (RKHS). 
A space $\Hilbert = \{ f : \X \to \Real \}$ of real-valued functions over $\X$ is a RKHS if any linear functional $L_x: \Hilbert \to \Real \,, \hspace{4pt} L_x(f) = f(x)$ is bounded in $\Hilbert$ (meaning that if two functions are close in terms of norm, then they are close also pointwise); or equivalently, for the Riesz representation theorem, it holds that $\forall x \in \X$ exists a unique $K_x \in \Hilbert$ such that $f(x) = L_x(f) = \langle f, K_x \rangle_\Hilbert \,, \hspace{4pt} \forall f \in \Hilbert$. 
For every RKHS $\Hilbert$ there is a unique $K$ such that $K(x, x') = \langle K_x, K_{x'} \rangle$, namely $K$ is a reproducing kernel for $\Hilbert$, and viceversa given a $K$ positive definite kernel there is a unique Hilbert space of functions on $\X$ for which $K$ is a reproducing kernel \cite{aronszajn1950theory}. The mapping $\phi: \X \to \Hilbert$ encoding a data point within the RKHS is a \emph{feature map}. Since 
\begin{equation}\label{eq:hilbertkernel}
    K(x, x') = \langle K_x, K_{x'} \rangle = \langle \phi(x), \phi(x') \rangle
\end{equation}
we can interpret the application of $K$ as calculating the inner product over a different vector space than the original point space $\X$. 

In supervised learning applications is common to use a feature map $\phi$ to encode the data in higher dimensional (Hilbert) space to find a linear separation of the transformed data and by using the inverse map $\phi^{-1}$ we can recover a complex, nonlinear decision boundary in the original space. Due to the \emph{representer theorem}, the linear pattern can be found independently of the dimensionality of $\Hilbert$: given the data points $\{ (x_1, y_1), ..., (x_m, y_m) \}$, the algorithms are fed with the kernel Gram matrix $K_{i,j} = [K(x_i, x_j)]$ of pairwise kernel similarities and no other information about the data is necessary for the classifier. Formally, the representer theorem asserts the linear function 
\begin{equation}
    \min_{f \in \Hilbert} L(f) + \lambda \lVert f \rVert
\end{equation}
that minimizes the empirical risk is always in the form:
\begin{equation}
    f(x) = \sum_{i=1}^m \alpha_i K(x, x_i).
\end{equation}
The terms $L$ is the loss function, e.g. the mean square error $L(f) = \frac{1}{m} \sum_{i=1}^m \lVert y - y' \rVert^2$. The term $\lambda \lVert f \rVert$, $\lambda > 0$ has regularization purposes, \textit{i.e.} penalizes high norm solutions thus preferring smooth functions over non-smooth ones. The determination of the $\{ \alpha_i \}$ values is a convex (efficient) optimization problem. 

Kernel methods can be applied to supervised learning tasks using the kernel ridge regression algorithm \cite{murphy2012machine}, which is a straightforward generalization of linear regression, and the support vector machine (SVM) \cite{cortes1995support}, which finds the linear classifier that maximizes the margin (\textit{i.e.} the minimum distance between the data points and the boundary, on both sides). 
The SVM usually finds a sparse solution, \textit{i.e.} a classifier whose output depends only on a few dataset items named support vectors. Kernel methods can be applied also to unsupervised learning tasks. Kernel PCA \cite{scholkopf1997kernel} is the straightforward extension of the principal component analysis algorithm. It finds the components in the higher dimensional Hilbert space that have larger variance. Kernels can be applied to clustering techniques too, including the k-means algorithm \cite{macqueen1967classification}. 

Kernel function can be parameterized, \textit{i.e.} depending on one or more hyper-parameters that can be trained according to some loss function or chosen using a grid search. A different approach is the \emph{multiple} kernel learning \cite{bach2004multiple}, which consists of defining multiple, fixed kernels and learning the most effective linear combination of such kernels.

\subsection{Quantum Kernels implementation}

\begin{figure}[htbp]
    \centering
    \begin{subfigure}[b]{0.35\textwidth}
        \centering
        \includegraphics[width=\textwidth]{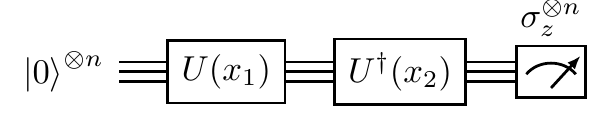}
        \caption{}
        \label{fig:overlap_test}
    \end{subfigure}
    \hfill
    \begin{subfigure}[b]{0.35\textwidth}
        \centering
        \includegraphics[width=\textwidth]{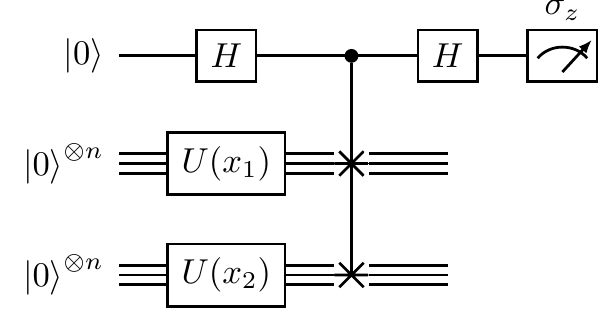}
        \caption{}
        \label{fig:cswap_test}
    \end{subfigure}
    \hfill
    \begin{subfigure}[b]{0.27\textwidth}
        \centering
        \includegraphics[width=\textwidth]{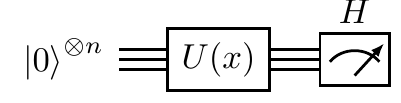}
        \caption{}
        \label{fig:proj_kernel}
    \end{subfigure}
    \caption{(\ref{fig:overlap_test})-(\ref{fig:cswap_test}) Fidelity test and SWAP test for Quantum Kernel Estimation where $U$ is the feature map associated with the quantum kernel. (\ref{fig:proj_kernel}) Quantum circuit for the feature map associated with the projected kernel, the Hermitian observable $H$ can be arbitrary.
    }
\end{figure}

Any parametric quantum circuit implementing the unitary transformation $U_\phi(\theta)$ acting on the Hilbert space $\Hilbert$ of the $n$-qubits quantum system can be used to implement a feature map:
\begin{align}
    & \phi: \X \to \Hilbert \nonumber \\
    & \ket{0 \cdots 0} \mapsto U_\phi(x) \ket{0 \cdots 0} = \ket{\phi(x)}
\end{align}
Such a feature map allows using the quantum space as an RKHS. In fact, we can obtain a kernel function sharing the same structure of Equation \ref{eq:hilbertkernel} by encoding a pair of data points into quantum states and calculating the inner product between them:
\begin{align}\label{eq:kernel}
    K(x, x') 
    & = \braket{\phi(x)}{\phi(x')} \\
    & = \text{prob}(\text{measurement of the state } U_\phi^\dagger(x')U_\phi(x)\ket{0\cdots 0} \nonumber \\
    & \phantom{= } \text{ using observable } \sigma_z^{\otimes n} \text{ collapse to eigenstate }\ket{0\cdots 0}).
\end{align}
Such a kernel can be concretely implemented using the overlap test circuit. The circuit structure is shown in Figure~\ref{fig:overlap_test}. 
We can equivalently use the SWAP test whose circuit structure is shown in Figure~\ref{fig:cswap_test}. 
Thus, we estimate the value of the kernel matrix $K_{i,j}$ by executing, for each pair of data points, $U_\phi$ multiple times (aka \textit{shots}). 
This procedure consists in performing multiple measurements which force the quantum wavefunction to collapse, resulting in the fidelity measure between the two encoded data points.
The kernel matrix can be finally fed to a kernel machine (e.g. SVM, Kernel PCA). Moreover, given the parametric quantum circuit for the feature map $U_\phi$ and a second quantum circuit $W$ implementing the state $\ket{w} = W\ket{0\cdots 0}$ corresponding to the linear weights, the function
\begin{equation}
    f(x) = \braket{\phi(x)}{w}
\end{equation}
is a linear classifier. 

Due to the large dimensionality of $\Hilbert$, exponentially in the number of qubits, the computation of the inner product may be affected by the \emph{curse of dimensionality}: any two pairs of quantum states uniformly sampled in the Hilbert space have a high probability to be almost orthogonal \cite{ball1997elementary}. 
Each off-diagonal element of the Gram matrix vanishes with the increasing dimension of $\Hilbert$.
If we think to perform the QKE on a current NISQ hardware this small value $K_{ij}$ becomes indistinguishable from the inherently pervading noise making our classifier worthless.
Such limitation requires a number of measurements to estimate a kernel function value that is polynomially in the dimensions of $\Hilbert$, thus exponentially in $n$.
Therefore, we need to accurately design our unitary transformation in order to avoid loosing quantum states within the Hilbert space. 

We can design an effective quantum transformation, \textit{i.e.} not affected by the curse of dimensionality, using several techniques. 

The first approach is the use of parametric quantum circuits that we know analytically are restricted to a small subspace of $\Hilbert' \subset \Hilbert$, \textit{i.e.} any parameter assignment $x$ results in $U(x)\ket{0\cdots 0} \in \Hilbert'$. 
The second approach is the use of a bandwidth coefficient \cite{canatar2022bandwidth}, \textit{i.e.} a small scalar to be applied pointwise to the components of $x$, diminishing the range of each component. 
The third approach is to implement a projected or biased quantum kernel \cite{Huang2021}, which projects the quantum state to an approximate classical representation through an observable $O$ (the choice of the observable is an educated guess).
The quantum state lives in a large Hilbert space, but the observable $O$ usually implies partial traces.
The effect of a partial trace (present in $O$) over, \textit{e.g.} the $k-th$ qubit, is to restrict the quantum space to some smaller representation, thus projecting it to the $k-th$ qubit subspace.
A projected kernel function could take a gaussian form as follows:
\begin{equation}
    k(x, x') = \exp(-\gamma \lVert \expval{U_\phi(x)OU_\phi^\dagger(x)}{0} - \expval{U_\phi(x')OU_\phi^\dagger(x')}{0} \rVert).
\end{equation}

The transformation $U_{\phi}$ defining the quantum kernel influences dramatically the linear decision boundary to be found in the feature space through classical optimization. In fact, we can find an optimal\footnote{A form which guarantees good accuracy and generalization, or that satisfies some objective function as the metrics we refer to further in this section.} form in an automatic fashion. 
The approach proposed by \cite{glick2021covariant} suggests the unitary transformation $U(x; \theta)$ should depend on both the data point features $x$ and on some trainable parameters $\theta$. 
The trainable parameters are then trained using stochastic gradient descent-based algorithm to minimize a loss function. 
Such an approach has been shown to be ineffective \cite{thanasilp2022exponential}. 
A different approach to optimize the parametric quantum circuit, choosing the basis gates at each point of the circuit as a combinatorial optimization algorithm (possibly a meta-heuristics) has been proposed by \cite{incudini2022structure, altares2021automatic}. 

To evaluate the performance of a quantum kernel, a family of metrics has been introduced: geometric difference, approximate dimension, model complexity, and target-kernel alignment.
The first three metrics constitute the central discussion of the Huang \textit{et al} paper \cite{Huang2021}.
The last one has several implementations already. The groundwork can be found in \cite{cristianini2001kernel}.

\begin{itemize}
\item The \textit{geometric difference} compares classical and quantum kernel feature spaces evaluating the separation in performances of the two kernels.
A large $g$ compared with the $\sqrt{N}$ indicates there is a deviation between the two kernel performances.

\item The \textit{approximate dimension} gives us an effective dimension of the quantum feature space generated by the encoding of the training samples.
Indeed, this quantity helps us to understand the expressibility of the quantum kernels.
If the $d$ saturates with $N$ it means the quantum states of the training data points are all orthogonal, otherwise a small value of $d$ tells us the Hilbert space has not been fully exploited and the model has limited expressivity.

\item The \textit{model complexity} represents a final test where we find the complexity of a kernel including in the computation the labels of data.
This metric derives from a prediction error generalization bound.

\item The \textit{target-kernel alignment}, as the model complexity, captures the relation present between a kernel and the relative target function, that is, the labels. 
The final objective of a kernel-based method is to approximate the label distribution with the data distribution in the feature space, and a margin tries to quantify this relation.
\end{itemize}

\section{Quantum Software Frameworks}\label{sec:software}

In recent years, a variety of software and programming languages have been developed to perform quantum computation.
Most of the frameworks express quantum computation in terms of quantum circuits \cite{feynman1985quantum}, which is the standard model \textit{de facto}. 
They are usually able to apply a universal set of basis gates, decompose a unitary matrix to a quantum circuit, reverse a circuit, perform uncomputation (e.g. to restore the original value of an auxiliary qubit), and perform circuit transformation (e.g. replacing part of a circuit with another one). 
Such frameworks allow the simulation on the host computer, while others allow sending the quantum circuit to some remote quantum hardware to be executed. 
Some possible alternatives to the quantum circuit model are the quantum lambda calculus \cite{van2004lambda}, the quantum Turing machine \cite{deutsch1985quantum}, the adiabatic quantum model \cite{farhi2000quantum}, the measurement-based quantum computation \cite{raussendorf2001one}, the topological quantum computation \cite{kitaev2003fault}, and the ZX calculus \cite{coecke2011interacting}. 

\begin{table}[tbp]
    \centering
\begin{tabular}{llllllll}
    \toprule
    Vendor & Name & Ref & Language & S & E & QML \\\midrule
    - & ProjectQ & \cite{steiger2018projectq} & Imperative & Yes & No & No \\
    - & QCL & \cite{omer2005classical} & Imperative & Yes & No & No \\
    - & QiBO & \cite{efthymiou2021qibo} & Python & Yes & No & No \\
    - & Quipper & \cite{green2013quipper} & Domain specific (Haskell) & Yes & No & No \\
    - & Quirk & \cite{quirk} & Drag-and-drop & Yes & No & No \\
    - & SilQ & \cite{silq} & Imperative & Yes & No & No \\
    Amazon & Braket & \cite{braket} & Python & Yes & Yes & No \\
    Baidu & Paddle Quantum & \cite{Paddlequantum} & Python & Yes & No & Yes \\
    Google & Cirq & \cite{cirq_developers_2022_6599601} & Python & Yes & Yes & No \\
    Google & TensorFlow Quantum & \cite{broughton2020tensorflow} & Python & Yes & Yes & Yes \\
    IBM & Qiskit & \cite{Qiskit} & Python & Yes & Yes & Yes \\
    Microsoft & Azure Quantum & \cite{AzureQuantum} & Python, Q\# & Yes & Yes & No \\ 
    Microsoft & LIQUi\textbar⟩ & \cite{wecker2014liqui} & Domain specific (F\#) & Yes & No & No \\
    Quantinuum & t\textbar{}ket⟩ & \cite{sivarajah2020t} & Python & Yes & Yes & No \\
    Rigetti & Forest (pyQuil) & \cite{pyquil} & Python & Yes & Yes & No \\ 
    Xanadu & PennyLane & \cite{bergholm2018pennylane} & Python & Yes & Yes & Yes \\
    Xanadu & Strawberry Field & \cite{strawberryfields} & Python & Yes & Yes & Yes \\  
    \bottomrule
\end{tabular}
    \caption{Comparison of relevant quantum computing frameworks. Legend: S: Simulation on CPU, E: Execution on quantum devices, QML: has quantum machine learning facilities. Domain-specific languages are embedded in the language denoted between parenthesis.}
    \label{tab:comparison}
\end{table}

A comparison of frameworks using the quantum circuit model is shown in Table~\ref{tab:comparison}. 
Most frameworks allow to import and export of circuits in the OpenQASM format \cite{cross2017open}, an open-source specification for quantum circuits. This facilitates the porting of quantum software among the different platforms.

\subsection{Quantum Machine Learning Frameworks}


PennyLane has been the first framework offering Quantum Machine Learning capabilities. 
They include the possibility to train a parametric quantum circuit, whose gradient can be calculated using the parameter-shift rule \cite{wierichs2022general} or with finite difference method. 
It allows the integration of a quantum transformation as a layer in a neural network object defined in Keras \cite{chollet2015keras} or PyTorch \cite{NEURIPS2019_9015} libraries. 
PennyLane has also facilities to define a quantum kernel, whose fidelity circuit (Figure \ref{fig:overlap_test}) is created automatically given the circuit for a quantum embedding. 
Strawberry Field proposes the same high-level capabilities for Continuous Variables formalism of quantum computing, like photonic Quantum Computing.

Qiskit Quantum Machine Learning has similar features, allowing us to embed quantum transformations within PyTorch networks and calculate kernel matrices.
TensorFlow Quantum \cite{broughton2020tensorflow} allows for rapid prototyping of hybrid-classical models due to its straightforward integration with the Machine Learning library TensorFlow \cite{tensorflow2015-whitepaper}. 
Paddle Quantum allows for effortless application of QNNs to define LOCC (Local Operations and Classical Communication) protocols \cite{chitambar2014everything}.
Moreover, it allows the simulation of some quantum machine learning algorithms defined in the measurement-based quantum computation formalism. 

\section{Proposed approach}

As described in Section \ref{sec:software}, many different quantum machine learning software exists, and most of them have few high-level algorithmic capabilities. 
However, there are several issues to address: many experiments require the interaction between different software platforms, e.g. Qiskit with PyTorch, requiring specific expert knowledge to be used. Furthermore, experiments need to be compared with theoretical results which are growing in the literature without, usually, a common implementation baseline. 

Therefore, we have designed QuASK, a unifying, easy-to-use software framework that automates each phase of an experiment: the selection of the dataset, the preprocessing, the definition of the kernel, its implementation, and analysis. 
QuASK can be used both as a standalone executable through its command line interface and as a software library. 
The first approach performs the experiment without writing a single line of code. 
The second approach might be interesting if the researcher needs to use both existing code routines. 
After having accurately processed the data, and implemented them to compute classical and quantum Gram matrices we have a modest range of metrics (proposed in QuASK) to evaluate the obtained kernel methods.

\subsection{Running experiments through a command line interface}\label{sec:script}

\begin{figure}[tbp]
    \centering
    \includegraphics[width=\textwidth]{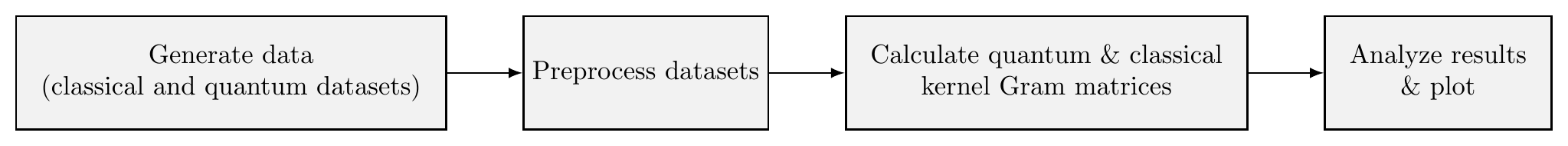}
    \caption{Sequence of operations performed when analyzing a dataset using QuASK.}
    \label{fig:pipeline}
\end{figure}

We show how to use QuASK to perform an end-to-end experiment. Once installed, the software is run with \texttt{quask <command>}\footnote{It can be equivalently run with \texttt{python3.x -m quask <command>} where \texttt{x} is the Python version installed on the researcher' system}. QuASK performs the sequence of operations illustrated in Figure \ref{fig:pipeline}. 

The experiment should start with the choice of a dataset. In such a case, QuASK offers several classical datasets both for regression and classification tasks. Moreover, some quantum datasets are available, i.e. datasets whose features has be encoded on a quantum system and modified by a unitary transformation, such as the one used in \cite{Huang2021}. The output of the process is a pair of NumPy binary files representing the feature data and the corresponding labels. \smallskip

\noindent
\includegraphics[width=1.0\textwidth]{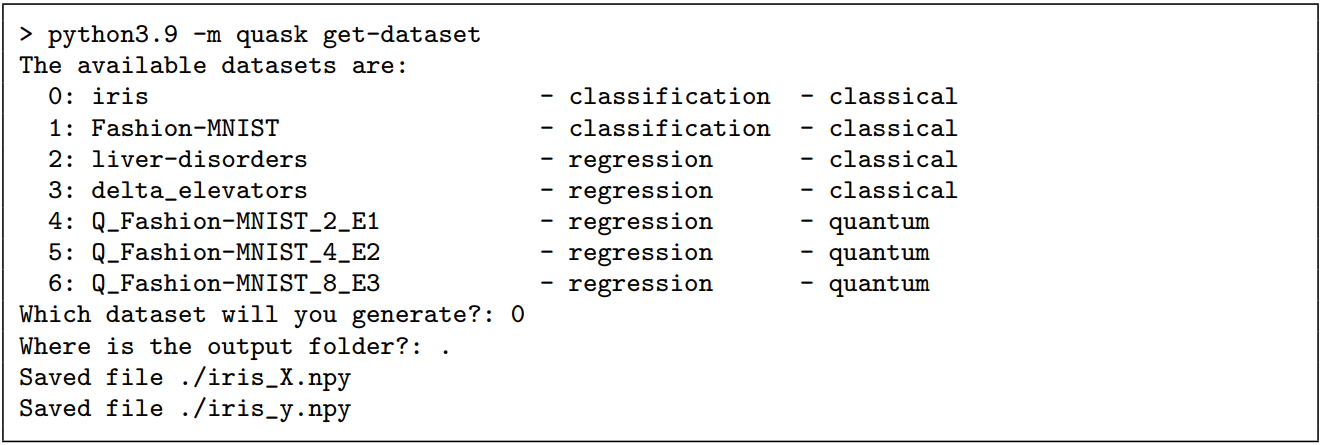}

\bigskip

The dataset, which can be obtained by using the QuASK command \texttt{get-command} or by using any dataset provided by the user in NumPy format (a feature matrix \texttt{X.npy}, $X \in \Real^{d \times n}$, and a label vector \texttt{y.npy}, $y \in \Real^{1 \times n}$), can be preprocessed classically before being fed to the quantum machine learning algorithm. Several preprocessing techniques are available. Firstly, the researcher can vertically slice the dataset, keeping only a certain range of labels. Specifically, the software prompts the researcher to simplify the classification task by restricting it to binary classification. However, it is worth noting that most kernel-based predictors are able to handle both binary and multi-class classification problems. Secondly, the user can apply dimensionality reduction techniques. These are important especially in the NISQ setting due to the lack of resources. The techniques available are PCA for numerical data and FAMD for mixed numerical and categorical data\footnote{FAMD, Factor Analysis for Mixed Data is implemented using Prince library (available at \url{https://github.com/MaxHalford/prince}).}. These choices are motivated by the fact that PCA is a widely used dimensionality reduction technique, while FAMD is a new method specifically tailored to handle categorical data. The user can extend QuASK to include further dimensionality reduction techniques. Thirdly, it is possible to fix the possible imbalanceness of the classes using random undersampling or random oversampling. When loading, the script already shows some statistics about the dataset, both for classification and regression tasks, which can guide the user through the preprocessing. The output of the process is the four files \texttt{X\_train}, \texttt{y\_train}, \texttt{X\_test}, \texttt{y\_test} which can now be fed to some kernel machine. \smallskip

\noindent
\includegraphics[width=1.0\textwidth]{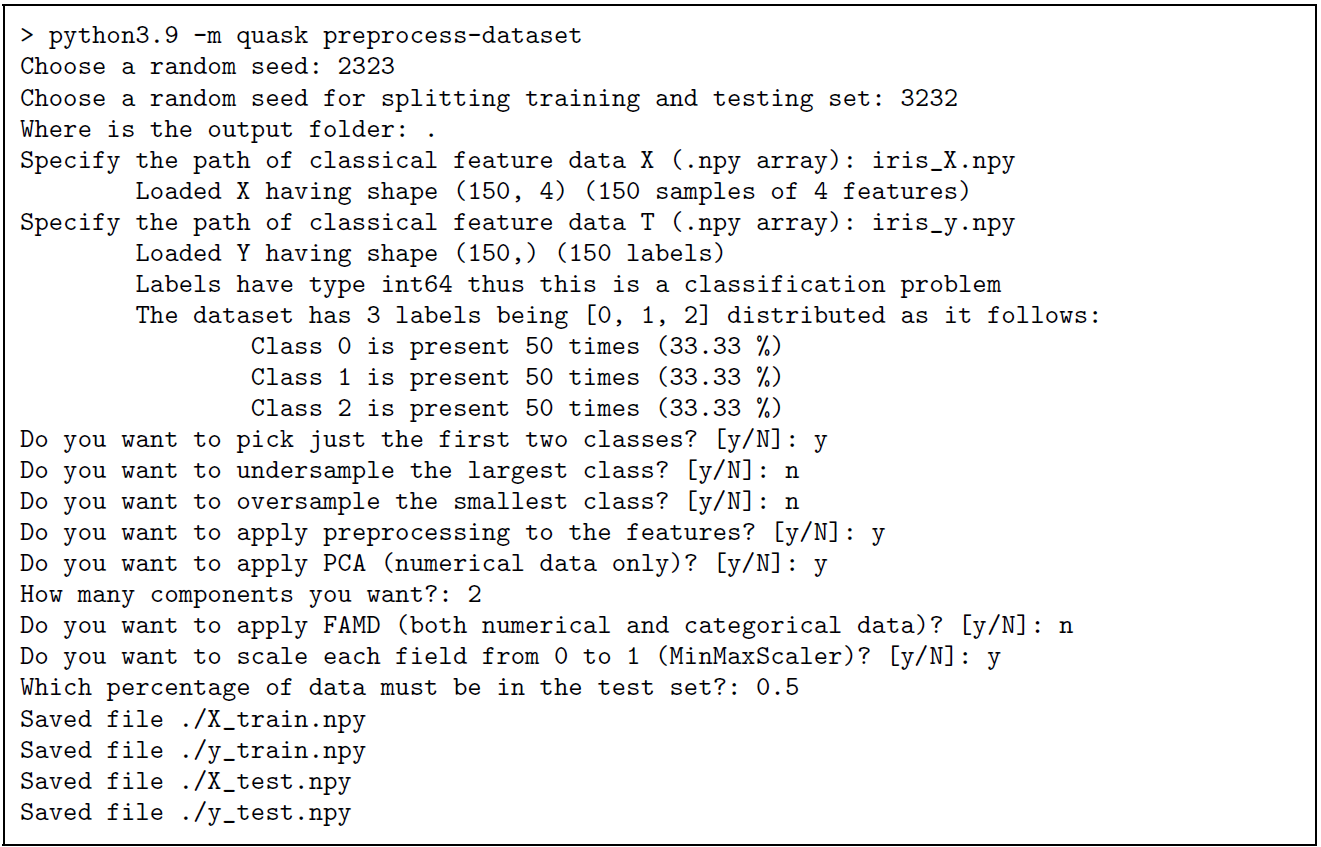}
\bigskip

At this point, the quantum kernel is built on the processed dataset. There are several available techniques the researcher can select from. The results of such an evaluation are the kernel Gram matrices corresponding to the training and testing datasets.
\smallskip

\noindent
\includegraphics[width=1.0\textwidth]{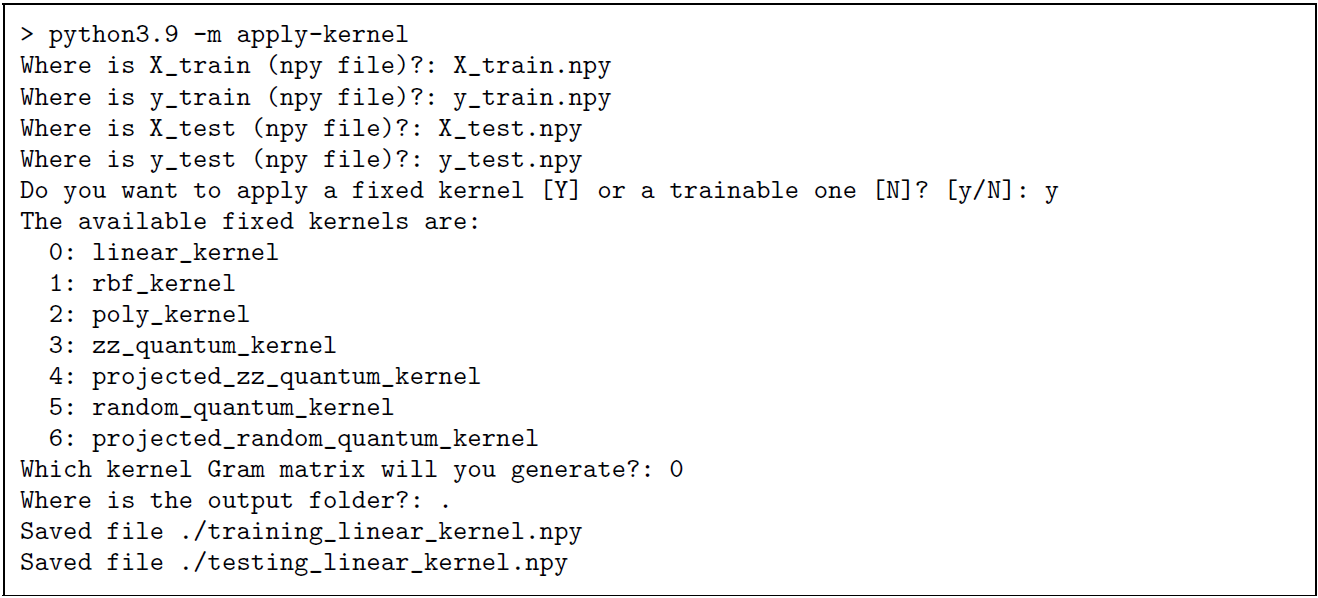}
\bigskip

The researcher can use \emph{optimized quantum kernels}, \emph{i.e.} quantum kernels whose circuits have been chosen after an optimization process. Such a process can be gradient-descent (ADAM optimizer) or gradient-free (grid search optimizer) based, in case we are optimizing the angles of the quantum operations, or combinatorial-optimization based, in case we are optimizing the generators of the quantum transformations. Although some quantum machine learning frameworks, such as PennyLane and Qiskit, already allows gradient-descent optimization of any circuit (including quantum kernels), no one offers the capabilities to adaptively choose the generators of the transformation through combinatorial optimization. \smallskip

\noindent
\includegraphics[width=1.0\textwidth]{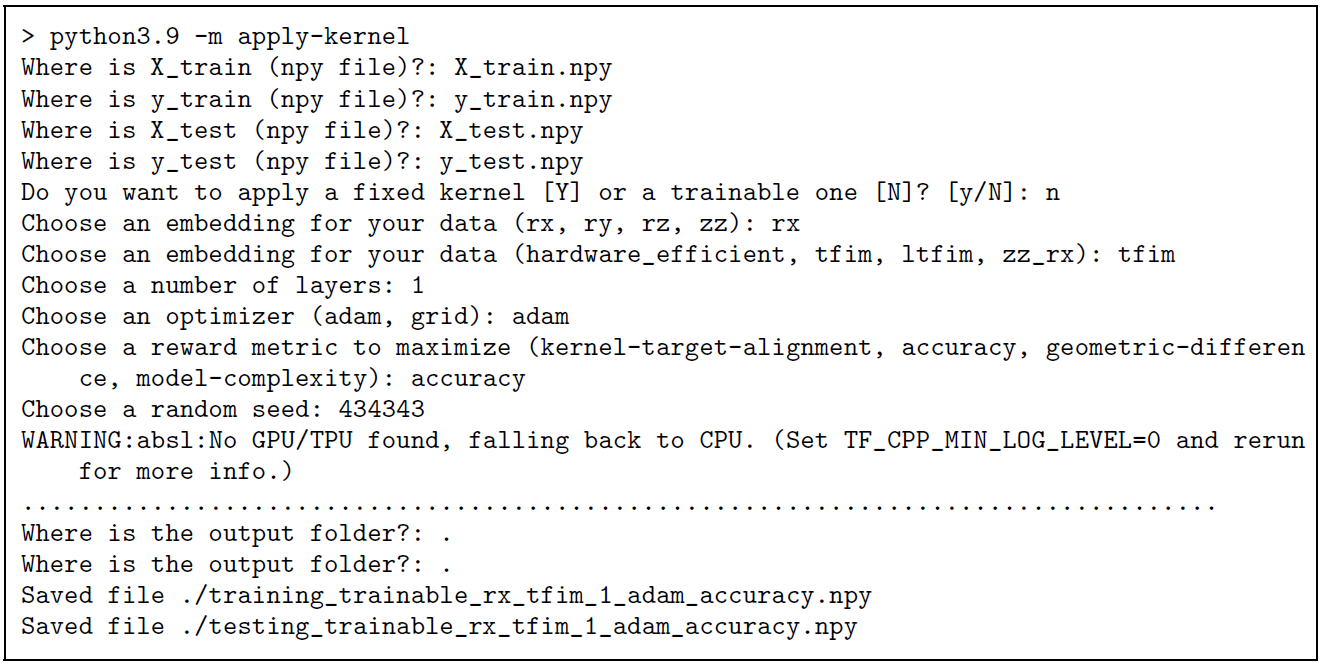}
\bigskip

Finally, the researcher can calculate the accuracy of the kernel model using the training and testing Gram matrices given as input. The output is a plot comparing the different kernels. For each kernel matrix, the user specifies the label that appears at the x-axis of the plot. If multiple instances are specified with the same label these are interpreted as i.i.d. random experiments and will contribute to the error bars. Multiple metrics are defined. 

\noindent
\includegraphics[width=1.0\textwidth]{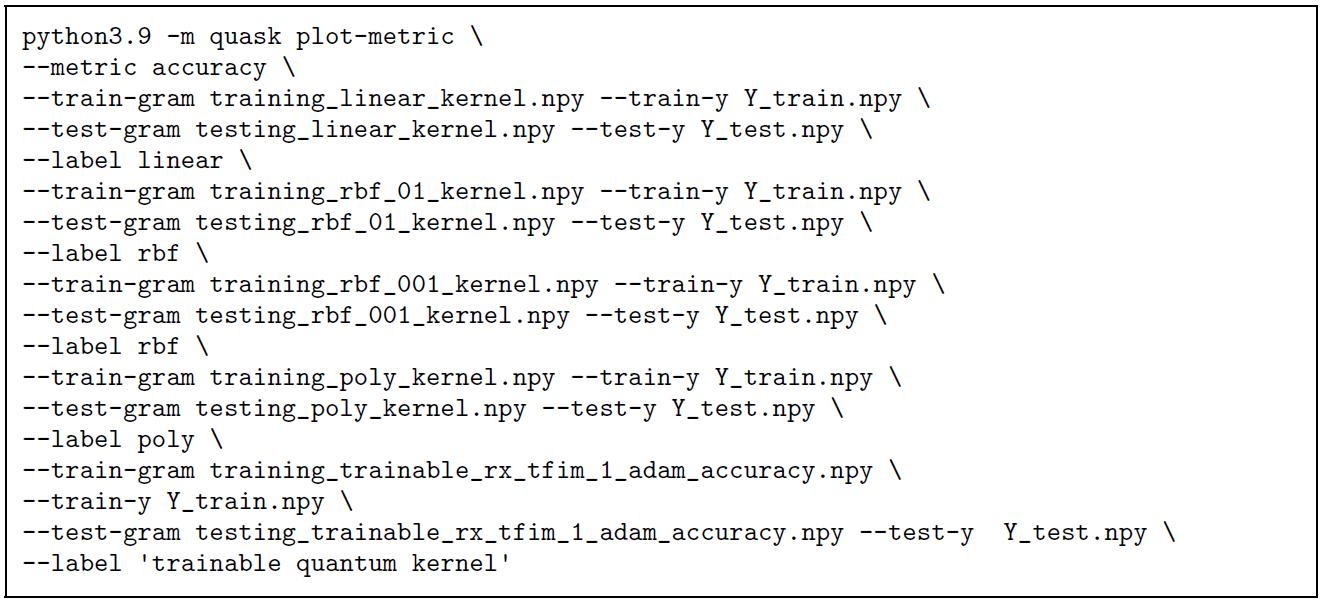}
\bigskip

\subsection{Integrating QuASK in an existing code base}

\begin{figure}[tbp]
    \centering
    \includegraphics[width=0.5\textwidth]{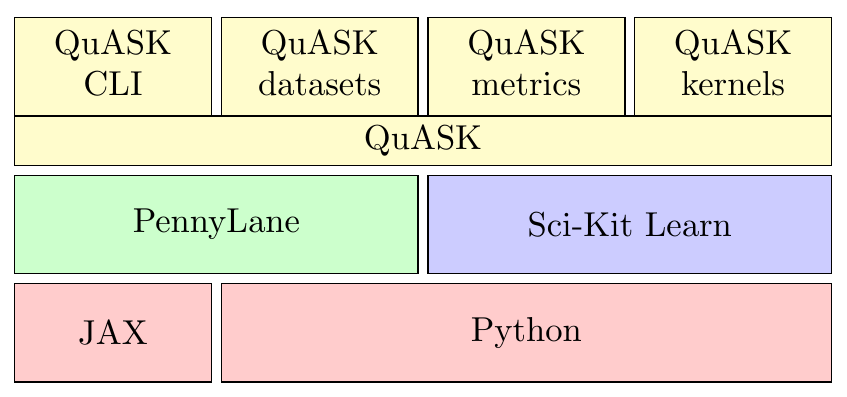}
    \caption{Software stack describing the modularity of QuASK. QuASK is written on top of PennyLane for defining the quantum circuit software and Sci-Kit Learn. These frameworks, written in Python, allow accessing basic machine-learning routines. }
    \label{fig:software-stack}
\end{figure}

There might be cases in which the command line interface of QuASK cannot be straightforwardly used for a certain project. For example, the researcher might be forced to use a particular preprocessing technique, or analyze the result accordingly to a custom metric. Such cases can be addressed by integrating QuASK with the existing code base. In fact, QuASK provides a library of elements that can be integrated with other projects. The software is organized into several modules whose structure is shown in Figure \ref{fig:software-stack}. 
  
\subsubsection{Download or generate datasets} 

The \texttt{quask.datasets} module facilitates the researcher in choosing a suitable dataset, providing some of the most popular datasets from OpenML platform\footnote{OpenML is an open platform for sharing datasets, available at https://www.openml.org/} and custom datasets generated also from quantum experiments (the latter allows us to reproduce results in \cite{Huang2021}). 

%
\noindent
\includegraphics[width=1.0\textwidth]{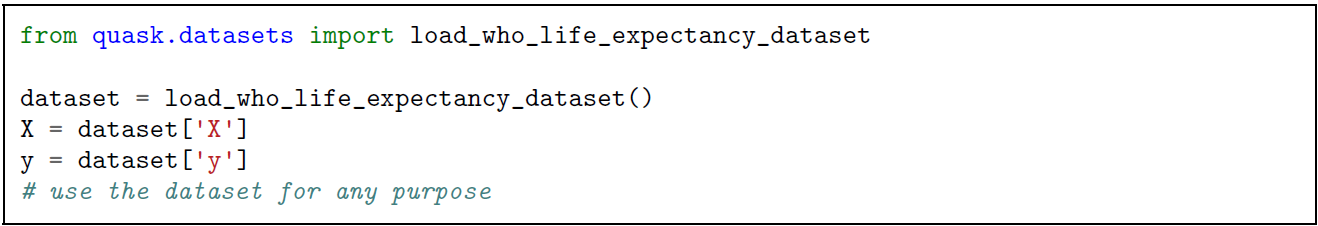}
\bigskip

\subsubsection{Evaluation metrics}

The quality of a quantum kernel can be empirically tested through the performance of a kernel machine with respect to a certain dataset, by evaluating some metrics. 
The module \texttt{quask.metrics} contains the metrics to compare and evaluate the kernels, including the kernel polarity (\emph{i.e.} Frobenius inner product between two Gram matrices), the target-kernel alignment \cite{cristianini2001kernel}, the training and testing accuracy of the Support Vector Machine with the precomputed kernel, the geometric difference \cite{Huang2021} (which can be used to find a potential quantum advantage), and the model complexity \cite{Huang2021}. 

%
\noindent
\includegraphics[width=1.0\textwidth]{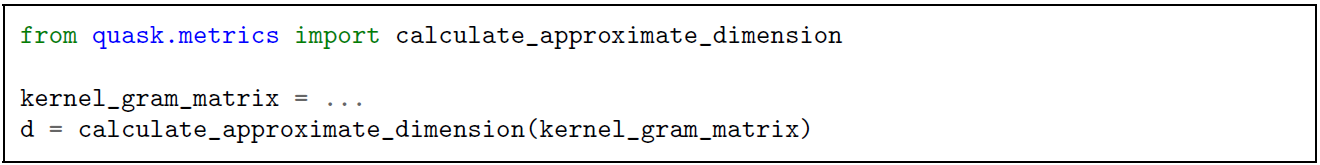}
\bigskip

\subsubsection{Implement quantum kernels}

The quantum kernel requires the definition of a feature map $\phi$ which is implemented using a parameterized unitary transformation $U_\phi(x)$. The Quantum Kernel Estimation algorithm \cite{havlivcek2019supervised} calculating the function $k(x, x') = \braket{\phi(x)}{\phi(x')}$ is implemented through either the \emph{overlap test} (Figure~\ref{fig:overlap_test}) or the \emph{C-SWAP test} (Figure~\ref{fig:cswap_test}). The projected quantum kernel calculates classically the inner product between two feature vectors $\phi(x_1), \phi(x_2)$, each one being the output of the quantum feature map $\phi(x) = \expval{U^\dagger(x) H U(x)}{0}$ (Figure~\ref{fig:proj_kernel}). The feature map crosses the quantum space first through U(x) and projects the data back in a classical representation when measuring with the Hermitian operator $H$.
QuASK contains both some notable unitary transformations $U$ from the literature and the code to use such unitary transformations as a kernel function through one of the three methods described above. The user can define their own unitary transformation and immediately get the kernel function. QuASK is agnostic with respect to the software framework used to define, simulate and execute the quantum circuits: we have implemented some unitary transformations in PennyLane. This allows also the use of the different functionalities offered by the different frameworks. For example, noiseless simulation with PennyLane can be speeded up using \texttt{JAX}\footnote{JAX is a high-performance linear algebra library.}. The open-source nature of QuASK allows for easy integration of other quantum computing frameworks in this platform.

The module \texttt{quask.kernels} collect all the quantum kernels defined within the platform. Most of the quantum kernels available are parametric quantum transformations in the form of Equation \ref{eq:kernel}. Such a module can be straightforwardly extended to include user-specified quantum kernels.
However, we can also have a more expressive  quantum transformation, parameterized by both the user features and some trainable parameters, which can be adjusted using gradient-descent-based techniques to minimize some criteria. Such criteria can be one of the functions implemented in the \texttt{quask.metrics} module. The user can take advantage of the efficient \texttt{optax} optimization library. Moreover, we can use Structure Learning techniques \cite{incudini2022structure} to optimize the generators of the transformation using a combinatorial-optimization-based technique such as Simulated Annealing \cite{kirkpatrick1983optimization} or Genetic Algorithms \cite{forrest1996genetic}. 

%
\noindent
\includegraphics[width=1.0\textwidth]{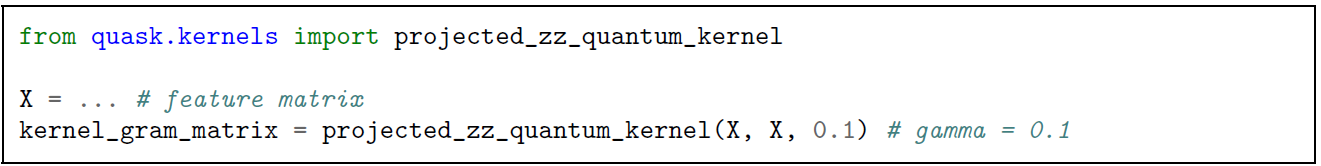}
\bigskip

\subsection{Execution on real hardware}

As the software is built on top of PennyLane, QuASK offers the same possibility of execution on real-world hardware. In particular, with the Qiskit-PennyLane plugin\footnote{\url{https://docs.pennylane.ai/projects/qiskit/en/latest}} it is possible to run the quantum circuit on the IBM superconductor-based quantum hardware, and with the Braket-PennyLane plugin\footnote{https://amazon-braket-pennylane-plugin-python.readthedocs.io/en/latest} it is possible to exploit Rigetti, IonQ, and Oxford Quantum Circuit hardware. The execution on the NISQ hardware is noisy and the results may largely deviate from the simulated ones. The authors in \cite{heyraud2022noisy} have studied the effect of noise on quantum kernels. 

\section{Conclusions}

We have introduced QuASK, a tool supporting researchers in creating powerful quantum kernels. The software takes care of the most time-consuming and error-prone aspects of the experimentation. It exploits theoretical metrics in QML, providing users with an environment to easily assess cases for potential quantum advantage. This package offers the exciting perspective of testing these metrics on real-world datasets. The QuASK project will be extended in future versions with a wider range of datasets and feature maps, both classical and quantum. 

\section*{Declarations}

\subsection*{Ethics approval and consent to participate}

Not applicable.

\subsection*{Consent for publication}

Not applicable. 

\subsection*{Availability of data and materials}

The software is freely available at \url{https://github.com/CERN-IT-INNOVATION/QuASK}. The documentation is available at \url{https://quask.readthedocs.io/en/latest/index.html}.

\subsection*{Competing interests}

Massimiliano Incudini and Davide Tezza are employees of Data Reply.

\subsection*{Funding}

Not applicable.

\subsection*{Authors' contributions}

All authors have contributed to the development of QuASK platform, and wrote the manuscript.

\subsection*{Acknowledgements}

Not applicable.



\end{document}